\documentclass[preprintnumbers,twocolumn,nofootinbib,amsmath,amssymb,aps,nofootinbib]{revtex4-1} 
\usepackage{graphicx}
\usepackage{dcolumn}
\usepackage[utf8]{inputenc}
\usepackage{bm}
\usepackage[caption=false]{subfig}
\usepackage{amssymb}
\usepackage{float}
\usepackage{hyperref}
\usepackage[T1]{fontenc} 
\usepackage{subfig}
\usepackage{dsfont}
\usepackage{slashed}
\usepackage{color}
\usepackage{amsmath}
\usepackage{mathtools}
\usepackage[section]{placeins}
\usepackage{braket}
\usepackage{upgreek}
\usepackage[bottom]{footmisc}
\usepackage[normalem]{ulem}
\usepackage{lipsum}
\usepackage[utf8]{inputenc}

\def\bs{\boldsymbol}

\def\med{{\rm med}}

\def\vac{{\rm vac}}

\def\bfq{{\bs q}}

\begin{document}

\title{Probing jet medium interaction with generalized projected energy correlators}

\author{Balbeer Singh}
\email{balbeers@lip.pt}
\affiliation{LIP, Av. Prof. Gama Pinto, 2, P-1649-003 Lisboa, Portugal }

\begin{abstract}
We study medium modifications to generalized projected energy correlators, referred to as $\nu$-correlators, measured on jets in heavy-ion collisions. While the vacuum distributions of these correlators exhibit a characteristic $1/\nu$ enhancement, we show that in a dense medium this behavior is regulated within both the BDMPS-Z and semi-classical approximations as $\nu \to 0$. As a result, contributions from medium-induced emissions become parametrically suppressed relative to the vacuum baseline. We observe a similar trend in the simulated events generated with JEWEL.  We find that while the large angle enhancement is qualitatively similar across different values of $\nu$, the small angle distribution changes drastically between $\nu<1$ and $\nu>1$. This indicates that small angle modifications, primarily driven by jet energy loss, are imprinted differently across $\nu$-correlators. In particular, for $\nu<1$ the small angle enhancement is less prominent which suggests that these correlators may provide improved sensitivity to medium modifications in the large angle region.
\end{abstract}

\maketitle

\section{\label{sec:introduction}Introduction}
Jets are collimated sprays of hadrons produced through the fragmentation and hadronization of energetic quarks and gluons (collectively referred to as partons) generated in hard-scattering events during the initial stages of high-energy particle collisions~\cite{Catani:2001cc,Salam:2010nqg}. Their evolution spans a wide range of energy scales, from the hard production scale, typically $\sim \mathcal{O}(100)$ GeV, down to the non-perturbative hadronization scale of $\sim \mathcal{O}(1)$ GeV, making jets a unique probe of both perturbative and non-perturbative aspects of quantum chromodynamics (QCD). Consequently, jet measurements in proton-proton (pp) collisions enable precision tests of QCD dynamics and provide important constraints on parton distribution functions~\cite{CMS:2024mlf,Dulat:2015mca}.

In heavy-ion collisions (HICs), jets undergo additional interactions, primarily through scattering with the constituents of the quark-gluon plasma (QGP). These scatterings drive the energetic color-charged jet parton off-shell, triggering gluon radiation known as medium-induced emission, which contributes to the measurements performed on final-state particles~\cite{Connors:2017ptx,Wiedemann:2000za,Blaizot:2012fh,Casalderrey-Solana:2011ule,Andres:2020vxs,Schlichting:2020lef,Guo:2000nz,Mehtar-Tani:2024mvl,Mehtar-Tani:2013pia,Salgado:2003rv,Mehtar-Tani:2025rty}. One important mechanism contributing to medium-induced radiation is the QCD analog of the Landau--Pomeranchuk--Migdal (LPM) effect, which arises from quantum interference between multiple scatterings of the jet parton with thermal partons in a dense medium~\cite{Landau:1953um,Migdal:1956tc,Gyulassy:2003mc}. A second contribution arises from quantum interference among multiple fast-moving color-charged partons within the jet itself, characterized by a coherence angle~\cite{Mehtar-Tani:2011lic,Casalderrey-Solana:2012evi,Mehtar-Tani:2017ypq,Mehtar-Tani:2017web}. Beyond these two mechanisms, hadrons sourced by energy-momentum transfer from the jet to the medium, known as medium response, can also contribute significantly to the experimentally measured final state particles; see~\cite{Cao:2020wlm} and references therein. Disentangling and probing these distinct features of jet-medium interaction dynamics remains one of the major ongoing efforts in both experimental and theoretical heavy-ion physics.

Over recent years, energy correlators, which measure angular correlations between jet constituents weighted by their energies, have emerged as promising jet observables that provide quantitative access to the multi-scale dynamics of medium-induced jet evolution and medium response in HICs~\cite{Yang:2026bsk,Barata:2026pgh,Barata:2025zku,Ke:2025ibt,Andres:2025yls,Alipour-fard:2024szj,Apolinario:2025vtx,Andres:2024xvk}. For a recent review on the applications of energy correlators for other collision systems see Ref.~\cite{Moult:2025nhu}. In the standard two point energy-energy correlator (EEC), it has been shown that medium-induced emissions and energy loss manifest as an enhancement in the large and small angle region of the correlator distribution relative to pp collisions~\cite{Andres:2023xwr,Andres:2022ovj,Andres:2024ksi}. In addition, medium response has been studied in Refs.~\cite{Yang:2023dwc,Bossi:2024qho,Barata:2024ukm}, and is found to contribute to similar large angular regions as medium-induced emissions. Experimentally, the first measurement of EECs in Pb-Pb collision has been performed by CMS collaboration~\cite{CMS:2025ydi}.

The two point energy correlator can also be generalized to N-point projected energy correlator, by tracking the largest angular separation between the $N$ final-state particles while integrating over the other angles.~\footnote{Recently, some efforts towards studying multi-point correlators in the medium have also been made. These correlators could also offer as a more differential probe of the medium-induced jet dynamics~\cite{Yang:2023dwc,Bossi:2024qho,Barata:2025fzd}. } Furthermore, analytic continuation in $N$ allows the projected energy correlator to be extended beyond integer values, so that it can be evaluated for non-integer values of $N$~\cite{Chen:2020vvp}. We denote this generalized index by $\nu$, and refer to these as $\nu$-correlators. A first study of $\nu$-correlators in HICs was performed in Ref.~\cite{Budhraja:2025ulx}, using a factorized approach within the effective field theory framework developed in Refs.~\cite{Mehtar-Tani:2024smp,Singh:2024pwr,Mehtar-Tani:2025xxd,Singh:2025scb,Singh:2025scb}. In the dilute-medium (single-scattering) limit, it was found that projected energy correlators with $\nu<1$ and $\nu>1$ exhibit distinct angular scaling behavior already at leading order.

In this article, we extend $\nu$-correlator study to a dense medium with multiple scattering scenario and compute $\nu$-correlator distributions using two well motivated distinct theoretical approaches; BDMPS-Z (i.e., soft gluon emission) and semi-classical approximation (i.e., hard gluon emission). At leading-logarithmic accuracy, we find that while the vacuum distribution exhibits a $1/\nu$ enhancement for $\nu<1$, medium-induced distributions do not display such behavior and instead approach a constant value as $\nu\to 0$. As a result, medium-induced distributions are suppressed relative to the vacuum baseline across all angular regions for small values of $\nu$, within both the multiple scattering approximations. We further observe a similar behavior in the events simulated with JEWEL Monte-Carlo generator~\cite{Zapp:2008gi,Zapp:2012ak,Zapp:2013vla}.   

Additionally, we observe that while the large angle behavior of the distributions remain qualitatively similar across $\nu<1$ and $\nu>1$, the small angle region changes drastically as we go from small to large $\nu$. This indicates that energy loss effects appear differently for $\nu<1$ than compared to $\nu>1$ values.  This represents a notable advantage, as it allows the relative contributions of different jet-medium interaction mechanisms to be tuned via $\nu$, offering a means to disentangle them. Therefore, a scan over a broad class of projected energy correlators, including values with $\nu<1$, offers new opportunities to isolate distinct medium contributions.

The rest of the article is organized as follows. In Sec.~\ref{sec:nucorr}, we briefly introduce $\nu$-correlators and derive the corresponding distributions at leading logarithmic accuracy in vacuum (pp collisions) in Sec.~\ref{sec:vacdist}. In Sec.~\ref{sec:meddist}, we extend the discussion to medium-induced distributions, considering both vacuum-like and medium-induced emissions while also accounting the effect of multiple scatterings with the medium. In Sec.~\ref{sec:jewelresults}, we compare the vacuum and medium-induced distributions for specific choice of $\nu$ values, namely $\nu=0.5,2,3$, for the events generated from JEWEL. Finally, in Sec.~\ref{sec:summary}, we summarize our results and discuss future directions.

\section{$\nu$-correlators~\label{sec:nucorr}}
Projected energy correlators (or $\nu$-correlators) are defined via analytic continuation of the (integer) $N$-point projected energy correlators, which generalize the standard two point energy correlator to $N$ final-state particles by considering the pair with the largest angular separation amongst them. This allows us to formally replace $N\to\nu$, with $\nu>0$ required for infrared safety. This analytic continuation places all $\nu$-correlators within a single family of observables and enables a single measurement to probe an infinite number of correlations, which can be made explicit by performing a binomial expansion of the weight function. For instance, for the two point case, the binomial expansion of the weights truncate to a finite number of terms, representing the self- and two-particle correlations. 

Formally, $\nu$-correlators are defined as~\cite{Chen:2020vvp,Budhraja:2024tev}

\begin{widetext}
\begin{align}
\frac{{\rm d}\Sigma^{[\nu]}}{{\rm d} \chi} = &\sum_M \int {\rm d}\sigma_X \bigg[\sum_{1\leq a_1 \leq M} {\cal W}_1^{[\nu]}(a_1)\,\delta(\chi)\, +  \sum_{1\leq a_1 < a_2 \leq M} {\cal W}_2^{[\nu]}(a_1,a_2)\,\delta(\chi-\Delta R_{a_1,a_2}) \, + \dots + \nonumber \\
& \!\!\sum_{1\leq a_1 < .. < a_M = M}\hspace{-20pt} {\cal W}_M^{[\nu]}(a_1,..,a_M)\,\delta(\chi \!-\!{\rm max}\{\Delta R_{a_1,a_2}, .., \Delta R_{a_{M-1,M}}\}) \bigg] ,
\label{eq:PEnuC}
\end{align}  
\end{widetext}
where ${\rm d}\sigma_X$ denotes the production cross section for the final state $X$, and $\Delta R_{ij}^2 = \Delta\eta_{ij}^2 + \Delta\phi_{ij}^2$ is the relative distance between particles $i$ and $j$ in the rapidity-azimuth plane. The weights ${\cal W}_{1,2,\dots}$ appearing above account for correlations between subsets of particles in the jet, with ${\cal W}_1^{[\nu]}$ representing all one-particle correlations, ${\cal W}_2^{[\nu]}$ representing all two-particle correlations, and so on, while $M$ denotes the total number of particles in the jet. 

For a single particle final state, the weight function simply reads as
\begin{equation}
\mathcal{W}^{[\nu]}_1(i_a)=\frac{E_{i_a}^{\nu}}{\omega^{\nu}} \equiv z_{i_a}^{\nu},    
\end{equation}
where $z_{i_a}$ is the energy fraction of particle $i_a$ inside the jet of radius $R$, and $\omega$ is the energy of the parton that initiates the jet. For a two-particle final state, the weight function is given by
\begin{equation}
\mathcal{W}^{[\nu]}_2(i_1,i_2)=\frac{(E_{i_1}+E_{i_2})^{\nu}}{\omega^{\nu}}-\sum_{a=1,2}\mathcal{W}^{[\nu]}_1(i_a).    
\end{equation}  
Weight functions for higher-multiplicity final states can be defined analogously. In this work, we restrict our attention to projected energy correlators for two-particle final states, for which the measurement function $\mathcal{M}^{[\nu]}$ is defined as 
\begin{align}
\mathcal{M}^{[\nu]}=\sum_{a=1,2}\mathcal{W}^{[\nu]}_1(i_a)\delta(\chi)+\sum_{i_1<i_2}\mathcal{W}^{[\nu]}_2(i_1,i_2)\delta(\chi-\theta_{i_1 i_2}) ,
\label{eq:meas}
\end{align}
where $\theta_{i,j}$ is the angle between the final state partons and $\chi$ is measurement. The first term represents the contact contribution arising from self-correlations of the partons, while the second term corresponds to correlations between two distinct partons.

\section{Vacuum emissions}
\label{sec:vacdist}
We begin by considering the vacuum case and define the differential cross section for $\nu$-correlators as
\begin{equation}
\frac{d\Sigma^{[\nu]}}{d\chi}=\sum_{i,j\in J}\int dz\, d\theta_{ij}\frac{1}{\sigma}\frac{d\mathcal{P}}{dz\,d\theta_{ij}}\mathcal{M}^{[\nu]}(z,\theta_{ij},\chi)  \end{equation}
where $\sigma$ is the total cross section and $\mathcal{M}^{[\nu]}$ is the measurement function for two-particle final states, defined in Eq.~\ref{eq:meas}. The sum over $i,j$ runs over particles inside the jet $J$, $z$ is the energy fraction of the emitted gluon, and $\theta_{ij}$ is the angular separation between particles $i$ and $j$. To gain qualitative insight into the $\nu$ dependence of the
correlators in HICs, we perform leading order analytical calculations for quark jets and use JEWEL simulations to study a more realistic phenomenological scenario. 

For the vacuum case, the splitting probability for $q\to q+g$ reads as
\begin{equation}
\frac{d\mathcal{P}_{\rm vac}}{d\theta\,dz}=\alpha_s(\mu)P_{qg}(z)\,\frac{1}{\theta} 
\label{eq:vac}
\end{equation}
where $\mu$ is the scale at which the coupling constant is evaluated. This scale is set by the logarithms appearing in the fixed-order calculation and is independent of $\nu$\footnote{Determining the appropriate scale $\mu$ that enters the coupling constant in energy correlators typically requires a dimensional-regularization treatment of the underlying phase-space integrals~\cite{Chen:2020vvp}.}. The leading-order splitting function $P_{qg}$ is given by
\begin{align}
P_{qg}=C_F\frac{1+(1-z)^2}{z},    
\end{align}
where $C_F=4/3$. We also consider cumulative distributions, which provide a more convenient framework for resummation. For this purpose, we define the cumulant, which at leading order reads as 
\begin{align}
\Sigma^{[\nu]}_{q}(\chi)&=\!\!=C_F\alpha_s\int_0^1 dz P_{qg}(z)\int_0^1 \frac{d\theta}{\theta}\Big[(z^{\nu}+(1-z)^{\nu})\Theta(\chi)\nonumber\\
&+(1-z^{\nu}-(1-z)^{\nu})\Theta(\chi-\theta)-\Theta(\chi) \Big]
\end{align}
where the first term proportional to $\Theta(\chi)$ correspond to self-correlations and originate from real gluon emission, while the last term is the virtual correction. To keep the expressions compact, we have omitted the scale dependence of the coupling constant, which we discuss later. Since we are interested in the dominant contribution, we evaluate the expression in the soft limit, i.e., $z\ll 1$. The virtual correction gets canceled with the second term in the first line. After performing all the integrations the distribution reads as
\begin{align}
\Sigma^{[\nu]}_{q}(\chi)&=C_F\alpha_s\log(\chi)\int_0^1dz\,P_{qg}(z)\,(z^{\nu}-\nu z)\nonumber\\
&=-C_F\alpha_s\log(\chi)\Big(\frac{1}{\nu}\Big)+\mathcal{O}(\nu)
\label{eq:vacdist}
\end{align}
where the corrections are of order $\mathcal{O}(\nu)$. In the limit $\nu\to 0$, the distribution scales as $\Sigma_q(\chi)\propto 1/\nu$, and consequently small-$\nu$ correlators are predominantly sensitive soft physics. Moreover, since this $1/\nu$ behavior emerges specifically in the soft limit, we will follow  this enhancement in the medium-induced contributions in the following sections. The differential distribution is obtained by differentiating Eq.~\ref{eq:vacdist} with respect to $\chi$.

The resummed cumulative distributions can be obtained systematically within the factorized framework by solving the DGLAP evolution equations. At leading-logarithmic accuracy, the resummed distribution for a quark jet reads as~\cite{Chen:2020vvp}
\begin{align}
\Sigma_q^{[\nu]}(\chi)=\Big(\frac{\alpha_s(\omega\chi)}{\alpha_s(\omega)} \Big)^{-\frac{\hat{\gamma}(1+\nu)}{\beta_0}}   
\label{eq:cumresm}
\end{align}
where $\beta_0=\frac{11}{3}N_C-\frac{4}{3}n_f T_f$ is the one-loop QCD beta function  with $N_c=3$, $n_f$ being number of flavors and $T_f=1/2$. $\hat{\gamma}^{(0)}(j)$ is the Mellin moment of the splitting function, defined as $\hat{\gamma}(j)=-\int_0^1 dx\, x^{j-1}P(x)$, with $P$ being the splitting function~\footnote{Here $j=1+\nu$.}. Explicitly, for a quark jet, $\hat{\gamma}(j)=\frac{\alpha_s}{4\pi}(\hat{\gamma}_{qq}^{(0)}(j)+\hat{\gamma}_{gq}^{(0)}(j))+\mathcal{O}(\alpha_s^2)$, where $\hat{\gamma}_{lm}^{(0)}(j)$ are defined as
\begin{align}
&\hat{\gamma}_{qq}^{(0)}(j)=-2C_F\Big(\frac{3}{2}+\frac{1}{j(j+1)}-2(\Psi(1+j)+\gamma_E) \Big)\nonumber\\
&\hat{\gamma}_{gq}^{(0)}(j)=-2 C_F\frac{2+j+j^2}{j(j^2-1)},
\label{eq:anomalous}
\end{align}
where $\Psi(j)$ is the digamma function and $\gamma_E$ is the Euler-Mascheroni constant. At the same accuracy, the coupling constant ratio in Eq.~\ref{eq:cumresm} can be expanded using the one-loop running coupling solution so that the cumulative distribution reduces to a simple power law in $\chi$, $\Sigma_q^{[\nu]}(\chi) \propto \chi^{\hat\gamma(1+\nu)}$. Differentiating with respect to $\chi$, the differential distribution scales as~\footnote{For very small $\nu$ values, the correlators become increasingly sensitive to non-perturbative effects. The leading non-perturbative contributions for $\nu<1$ have been studied in Ref.~\cite{Budhraja:2026pyi}, which will be important for phenomenological applications.}
\begin{equation}
\frac{d\Sigma^{[\nu]}_q}{d\chi}\equiv c \frac{\hat{\gamma}(1+\nu)}{\chi^{1-\hat{\gamma}(1+\nu)}}  
\end{equation}
where $c$ is a constant which is independent of $\nu$. As $\nu\to 0$, similar to the fixed-order expression obtained in Eq.~\ref{eq:vacdist}, the resummed distribution exhibits the same approximate $1/\nu$ scaling as the fixed-order result and is dominated by $\hat{\gamma}^{(0)}_{gq}$ (see Eq.~\ref{eq:anomalous}). The full resummed result  along with higher-order computations, can be found in Refs.~\cite{Chen:2020vvp}.  We now turn to medium-induced distributions, focusing on $\nu<1$ values in order to identify the angular regions in which medium effects are suppressed or enhanced. We will explicitly check the small $\nu$ behavior of the distributions same as that in vacuum case.

\section{In-medium emissions} 
\label{sec:meddist}
\begin{figure}[h]
\centering 
\includegraphics[width=0.7\linewidth]{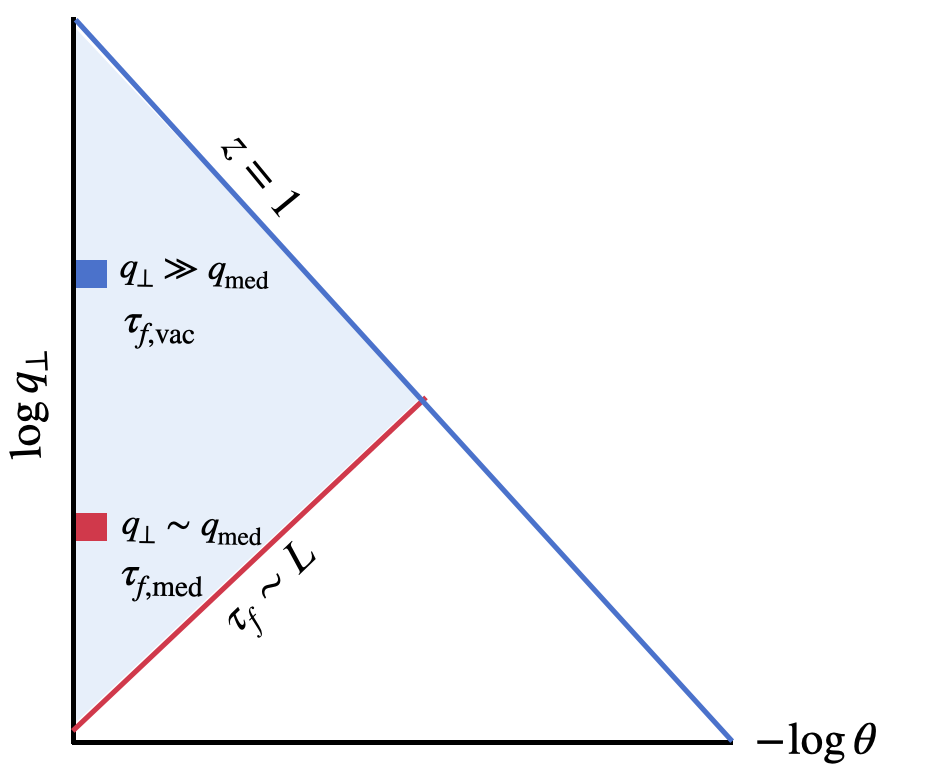}
\caption{Lund-plane representation of the phase-space regions associated with the formation times of medium-induced ($\tau_{f,\text{med}}$) and vacuum-like ($\tau_{f,\text{vac}}$) emissions as given in Eq.~\ref{eq:ftime}.} 
\label{fig:lund}
\end{figure}

For computing $\nu$-correlators in HICs we focus primarily on gluon emissions that contribute to medium-induced jet dynamics.\footnote{Note that medium response can also impact the distribution of $\nu$-correlators in HICs. However, the incorporation of these effects in beyond the scope of this work and will be addressed in a future study.  }   These in-medium emissions and the corresponding phase-space constraint are represented in the Lund plane in Fig.~\ref{fig:lund}. Here, the $y$-axis denotes the transverse momentum ($q_{\perp}=|\bfq|$) of the emitted gluon, the $x$-axis denotes its angle ($\theta)$ from the energetic quark, and similar to previous cases $z$ is the energy fraction carried by the emitted gluon. The blue diagonal line at $z=1$ represents hard gluon emissions and is obtained by using the relation $|\bfq|=z\theta \omega$, with $\omega$ being the energy of the incoming quark. The red line, defined by the formation time (of the emitted gluon) $\tau_f=z\omega/\bfq^2 = L$, where $L$ is the total length of the medium, marks the phase-space boundary separating gluon emissions formed inside versus outside the medium. The region to the right of this line corresponds to emissions with $\tau_f> L$, while the region to the left corresponds to $\tau_f< L$. Since emissions with $\tau_f>L$ are formed outside the medium, they do not contribute to the measurements performed in the final state particles.

In a dense medium, jet partons can undergo multiple scatterings with the medium partons. These multiple scatterings can act coherently over the formation time of the gluon emission. As a result, for a medium of length $L$ and $\tau_f\sim L$, the total transverse momentum of the emitted gluon saturates at $Q_s^2\equiv \hat{q}L$, where $\hat{q}$ is the jet quenching parameter. Moreover, at time $\tau_f$, the transverse momentum of the gluon emission is $q_{\rm med}=\sqrt{\hat{q}\tau_f}$. Emissions whose transverse momentum saturates at $Q_s$ are therefore referred to as medium-induced emissions. In Fig.~\ref{fig:lund}, the transverse momentum of these emissions is represented by the red rectangle at $|\bfq|=q_{\rm med}$, with formation time denoted by $\tau_{f,\med}$.

Emissions with transverse momentum $|\bfq|\gg q_{\rm med}$, on the other hand, can be generated through the vacuum shower of the initial hard parton that initiates the jet~\cite{CrispimRomao:2025yiq}. These are known as vacuum-like emissions, and we denote their formation time by $\tau_{f,\vac}$, shown with the blue rectangle in Fig.~\ref{fig:lund}. In general, emissions with large transverse momentum can also arise from a hard scattering between the jet and medium partons~\cite{DEramo:2018eoy}; however, we do not consider such emissions in this work. 

\subsection{vacuum-like emissions}

Since the transverse momentum of vacuum-like emissions is large compared to that of medium-induced emissions, i.e., $|\bfq|\gg q_{\rm med}$, their formation time is correspondingly much shorter, i.e., $\tau_{f,\med}\gg \tau_{f,\vac}$. The formation times of vacuum-like and medium-induced emissions are explicitly given as
\begin{equation}
\tau_{f,\vac}=\frac{z\omega}{\bfq^2}, \hspace{1cm}\, \tau_{f,\med}=\sqrt{\frac{z\omega}{\hat{q}}}, 
\label{eq:ftime}
\end{equation}
where $\tau_{\rm f, med}$ is obtained by replacing $\bfq^2$ with the momentum acquired in time $\tau_{\rm f, med}$ through medium-induced broadening, i.e., $\bfq^2 = \tau_{f,\med}\,\hat{q}$.

The formation time scale hierarchies reflect the fact that while both of them contribute to the measurement they populate the different region of the phase space (see Fig.~\ref{fig:lund}). Vacuum-like emissions are emissions are generated as early emissions and are  allowed in the region specified by $\Theta(\tau_{f,\med}-\tau_{f,\vac})\Theta(L-\tau_{f,\vac})$, where the second $\Theta$-function enforces that the emission is formed inside the medium. Therefore, for vacuum-like emissions, the differential distribution of the projected $\nu$-point energy correlator reads as
\begin{align}
&\frac{d\Sigma^{[\nu]}_{q,{\rm vle}}}{d\chi}= \sum_{i,j\in J}\int dz\, d\theta_{ij}\frac{d\mathcal{P}}{dz\,d\theta_{ij}}\mathcal{M}^{[\nu]}(z,\theta_{ij},\chi)\nonumber\\
&\quad \Theta(\tau_{f,\med}-\tau_{f,\vac})\Theta(L-\tau_{f,\vac}),
\label{eq:vle0}
\end{align}
where the indices $i,j$ denote particles inside the jet and $\theta_{ij}$ is the angle between the pair. Substituting the measurement function and splitting function defined in Eqs.~\ref{eq:meas} and \ref{eq:vac}, the leading order distribution takes the form
\begin{align}
&\frac{d\Sigma^{[\nu]}_{q,{\rm vle}}}{d\chi}=\frac{\alpha_s C_F}{\pi}\int dz\, P_{qg}(z) \int\frac{d^2\bfq}{\bfq^2}\Big[(z^{\nu}+(1-z)^{\nu})\delta(\chi)\nonumber\\
&\quad+(1-z^{\nu}-(1-z)^{\nu})\delta\Big(\chi^2-\frac{\bfq^2}{[z(1-z)\omega]^2}\Big)\Big]\nonumber\\
&\quad \Theta(\tau_{f,\med}-\tau_{f,\vac})\Theta(L-\tau_{f,\vac}).
\label{eq:vle}
\end{align}
 Note that for convenience, we have converted the angular integration into a transverse momentum integration. As mentioned earlier, to gain insight into the behavior of the distribution as $\nu\to 0$, we further simplify Eq.~\ref{eq:vle} in the soft limit, i.e., $z\ll 1$. Using Eq.~\ref{eq:ftime} for formation times, we obtain

\begin{align}
\frac{d\Sigma^{[\nu]}_{q,{\rm vle}}}{d\chi}=&\frac{\alpha_sC_F}{2 \pi}\int\frac{dz}{z}\int\frac{d\bfq^2}{\bfq^2}\Big[(z^{\nu}-\nu z)\delta(\chi)+(\nu z-z^{\nu})\nonumber\\
&\delta\Big(\chi^2-\frac{\bfq^2}{z^2\omega^2} \Big)\Big]\Theta(\bfq^4-z\omega\hat{q})\Theta(L\bfq^2-z\omega),
\label{eq:vle2}
\end{align}
where the term proportional to $\delta(\chi)$ is the contact term, which we drop from here onward. For projected energy correlators with $\nu>1$, the first term, i.e., $\nu z$, dominates the above equation, while for $\nu<1$ the second term gives the dominant contribution. Furthermore, the formation-time constraints on vacuum-like emissions set the limits of the $z$-integration. Performing the momentum integration using the measurement delta function, we get
\begin{align}
\frac{d\Sigma^{[\nu]}_{q,{\rm vle}}}{d\chi}&=\frac{\alpha_sC_F}{2 \pi\chi}\int_{z_{\rm min}}^1 dz (\nu-z^{\nu-1})\Theta(L\omega\chi^2 z-1)\nonumber\\
&\qquad\quad \times\Theta(z^3\omega^3\chi^4-\hat{q}),
\label{eq:vlesimp}
\end{align}
the lower limit of $z$ integral is set by the two theta functions which is given by
\begin{equation}
z_{\min}=\max\left\{\frac{1}{L\omega\chi^2},\ \left(\frac{\hat{q}}{\omega^3\chi^4}\right)^{1/3}\right\}.
\end{equation}
Carrying out the $z$-integration explicitly, we find
\begin{equation}
\frac{d\Sigma^{[\nu]}_{q,\rm vle}}{d\chi}=\frac{\alpha_sC_F}{2\pi\chi}\left[\nu\big(1-z_{\min}\big)-\frac{1}{\nu}+\frac{z_{\min}^{\nu}}{\nu}\right].
\label{eq:vle17}
\end{equation}
For small-$\nu$ limit of Eq.~\ref{eq:vle17}, we write $z_{\min}^\nu=e^{\nu\log z_{\min}}$ and expand in powers of $\nu$, which gives
\begin{equation}
\frac{d\Sigma^{[\nu]}_{q,\rm vle}}{d\chi}=\frac{\alpha_sC_F}{2\pi\chi}\left[\log z_{\min} + \nu\Big((1-z_{\min})+\frac{1}{2}\log^2 z_{\min}\Big)  \right]    
\end{equation}
where we have dropped the corrections are of order $\mathcal{O}(\nu^2)$ and higher.  From above expression we note that though vacuum-like emissions exhibit a similar fixed order power law structure in $\chi$ as Eq.~\eqref{eq:vacdist}, the phase-space restrictions corresponding to formation time lead to an exact cancellation of $1/\nu$ enhanced contributions, regardless of which $\Theta$-function sets the lower limit $z_{\min}$. Furthermore, different from the vacuum case, the vacuum-like emissions populate a specific angular regions of the correlator.  The crossover between the two regimes of phase space is set by the two theta functions appearing in the $z$-integration. Equating the two branches gives the crossover scale
\begin{equation}
\chi^{*}=\frac{1}{\sqrt{\hat{q}L^3}}.
\end{equation}
For $\chi<\chi^{*}$, the constraint $\Theta(Lz\omega\chi^2-1)$ dominates, giving $z_{\min}=1/(L\omega\chi^2)$, while for $\chi>\chi^{*}$, the constraint $\Theta(z^3\omega^3\chi^4-\hat{q})$ takes over leading to $z_{\min}=(\hat{q}/(\omega^3\chi^4))^{1/3}$.  For medium parameters $\hat{q}=1.5$~GeV$^2$fm$^{-1}$ and $L=5$~fm the cross-over scale is $\chi^{*}\approx 1.4\times 10^{-2}$. 

\subsection{Multiple scatterings: BDMPS-Z}

For medium-induced emissions resulting from multiple scatterings, we employ two approximations. First, we consider the BDMPS-Z spectrum~\cite{Baier:1996kr} which is valid in the limit where the emitted gluon is soft and its formation time lies in the region $\tau_f\leq L$. Second, we use semi-classical approximation~\cite{Dominguez:2019ges,Isaksen:2020npj}, which allows the emitted gluon to be hard and neglects the deflection from the classical trajectory, so that the evolution in the medium is governed by the rotation of the color fields. These two approximations allow us to qualitatively understand the small-$\nu$ behavior of generalized projected energy correlator distributions in two different kinematic regimes as well as their model dependence. As mentioned earlier, this is represented by the red rectangular region in Fig.~\ref{fig:lund}.

The splitting probability, including transverse-momentum broadening, for medium-induced gluon emissions in the BDMPS-Z approximation for the $q\to q+g$ process is given by
\begin{align}
\frac{d\mathcal{P}}{dz d\theta}&=\frac{\alpha_{s,\med}}{L}\sqrt{\frac{8\omega_c}{z^3\omega}}\Theta(\omega_c-z\omega)(z^2\omega^2\theta) \int_0^L\,\frac{dt\, e^{-\frac{z^2 \omega^2\theta^2}{\hat{q}(L-t)}}}{\hat{q}(L-t)}  
\label{eq:bdmpsz}
\end{align}
where $\omega_c=\hat{q}L^2/2$ is the maximum energy acquired by the emitted gluon at the phase-space boundary $\tau_f=L$, and the exponential term describes the broadening of medium-induced emission from its production point to the phase-space boundary, restricted by the size of the medium. The energy spectrum in Eq.~\ref{eq:bdmpsz} is valid in the region where the energy of the emitted gluon is smaller than $\omega_c$, i.e., $z\omega< \omega_c$, and $|\bfq|\ll Q_{s}\sim\sqrt{\hat{q}L}$. Furthermore, the coupling constant $\alpha_{s,\med}$ should formally be evaluated at the medium-induced transverse scale $q^2_{\rm med}\sim \hat{q}\tau_f$. Since our focus is on $\nu$-dependence rather than precise phenomenology, we treat it as a constant and set $\alpha_{s,\med}=0.3$ throughout. Using Eq.~\ref{eq:bdmpsz}, the medium contribution to $\nu$-correlator distribution reads as
\begin{align}
&\frac{d\Sigma^{[\nu]}_{q,\med}}{d\chi}=\frac{\alpha_{s,\med}}{\hat{q}L}\sqrt{2\omega_c \omega^5}\int dz (\nu z^{5/2}-z^{\nu+3/2}) \nonumber\\
&\Theta(\omega_c-z\omega)\int d\bfq^2\, \Gamma\Big(0,\frac{\bfq^2}{\hat{q}L}\Big)\delta(\chi^2 z^2\omega^2-\bfq^2)\Theta(Q_s-|\bfq|),
\label{eq:bdmpszdist}
\end{align}
where, to remain consistent with the approximations above, we have also taken the soft-gluon limit in the measurement function shown in Eq.~\ref{eq:bdmpszdist}. Here, $\Gamma(0,x)$ is the incomplete gamma function, defined as $\Gamma(a,x)=\int_x^{\infty} dt\, t^{a-1} e^{-t}$. In this limit, for projected energy correlators with $\nu>1$, the dominant contribution comes from the first term in the first line of Eq.~\ref{eq:bdmpszdist}, i.e., $\nu z^{5/2}\Gamma(0,x)$. On the othr hand, for small $\nu$ the final-state distribution is dominated by the second term, which can be seen by taking the $\nu\to 0$ limit and expanding it as $z^{\nu+3/2}\sim z^{3/2}+\nu z^{3/2} \log(z)$. This shows that the small-$\nu$ behavior of the distribution originates from $-z^{3/2}\Gamma(0,x)+\mathcal{O}(\nu)$. Note that this also confirms the small and large valued $\nu$-correlators scale differently with angle $\chi$. Furthermore, for $\nu=1$ the differential distribution vanishes, as anticipated from the measurement function in Eq.~\ref{eq:meas}.\footnote{In this case, only single-particle correlations are finite and contribute solely to the contact term.}

\begin{figure}[t]
\centering 
\includegraphics[width=1\linewidth]{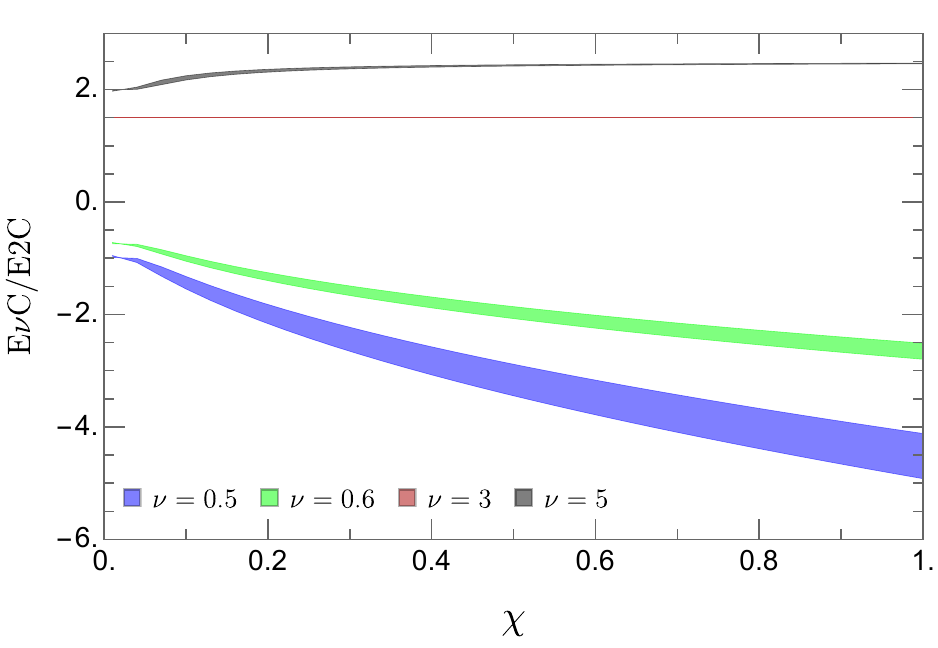}
\caption{Ratios of the medium-only $\nu$-correlators to the standard two point energy correlator for $\nu$ ranging from $0.5$ to $5$ as a function of the angular separation $\chi$. The distributions are obtained using the BDMPS-Z spectrum given in Eq.~\ref{eq:bdmpszdist}, for an initial quark energy $\omega=100$ GeV and jet quenching parameter $\hat{q}=[1-2]$ GeV$^2$fm$^{-1}$. The medium parameters are set to $L=5$ fm and $\alpha_{s,\med}=0.3$.} 
\label{fig:ratiobdmpsz}
\end{figure}

To obtain the analytic structure in the small $\nu$ regime, we simplify Eq.~\ref{eq:bdmpszdist} in $\nu\ll 1$ limits. To this end, we first perform the $\bfq^2$-integration using the delta function, which sets $\bfq^2=\chi^2z^2\omega^2$ and leaves
\begin{align}
\frac{d\Sigma^{[\nu]}_{q,\med}}{d\chi}&=\frac{\alpha_{s,\med}}{\hat{q}L}\sqrt{2\omega_c\omega^5}\int_0^{z_{\max}} dz\,(\nu z^{5/2}-z^{\nu+3/2})\nonumber\\
&\qquad\times \Gamma\Big(0,\frac{\chi^2z^2\omega^2}{\hat{q}L}\Big),
\label{eq:bdmpszdist2}
\end{align}
where $z_{\max}=\min\{\omega_c/\omega,\,Q_s/(\chi\omega)\}$ from the remaining two theta functions.  Defining $a=\chi^2\omega^2/(\hat{q}L)$ and using the general integral $J_p(z)\equiv\int z^p\,\Gamma(0,az^2)\,dz$, integration by parts (using $\frac{d}{dz}\Gamma(0,az^2)=-2e^{-az^2}/z$) gives
\begin{align}
J_p(z)=\frac{z^{p+1}}{p+1}\,\Gamma(0,az^2)-\frac{1}{(p+1)\,a^{(p+1)/2}}\,\Gamma\Big(\frac{p+1}{2},az^2\Big).
\label{eq:Jp}
\end{align}
Moreover, using Eq.~\ref{eq:Jp} to the two terms in Eq.~\ref{eq:bdmpszdist2} with $p=5/2$ term reads as
\begin{align}
\int_0^{z_{\max}} z^{5/2}\Gamma(0,az^2)\,dz&=\frac{2}{7}z_{\max}^{7/2}\Gamma(0,az_{\max}^2)-\frac{2}{7\,a^{7/4}}\nonumber\\
&\times\Big[\Gamma\Big(\frac74,az_{\max}^2\Big)-\Gamma\Big(\frac74\Big)\Big],
\label{eq:p52}
\end{align}
while for $p=\nu+3/2$ we find
\begin{align}
&\int_0^{z_{\max}} z^{\nu+3/2}\Gamma(0,az^2)\,dz=\frac{z_{\max}^{\nu+5/2}}{\nu+\frac52}\Gamma(0,az_{\max}^2)\nonumber\\
& -\frac{1}{(\nu+\frac52)\,a^{\frac\nu2+\frac54}}\Big[\Gamma\Big(\frac\nu2+\frac54,az_{\max}^2\Big)-\Gamma\Big(\frac\nu2+\frac54\Big)\Big].
\label{eq:pnu}
\end{align}
With $z_{\max}=\omega_c/\omega$, the argument of the incomplete gamma functions becomes $az_{\max}^2=\chi^2\omega_c^2/(\hat{q}L)$. Combining all the terms from Eq.~\ref{eq:p52} and Eq.~\ref{eq:pnu}, the medium induced distribution reads as
\begin{align}
&\frac{d\Sigma^{[\nu]}_{q,\med}}{d\chi}=\frac{\alpha_{s,\med}}{\hat{q}L}\sqrt{2\omega_c\omega^5}\Bigg\{\nu\Bigg[\frac{2}{7}z_{\max}^{7/2}\Gamma(0,az_{\max}^2)-\frac{2}{7a^{7/4}}\nonumber\\
&\quad\times\Big(\Gamma\Big(\frac74,az_{\max}^2\Big)-\Gamma\Big(\frac74\Big)\Big)\Bigg]
-\Bigg[\frac{z_{\max}^{\nu+5/2}}{\nu+\frac52}\Gamma(0,az_{\max}^2)\nonumber\\
&-\frac{1}{(\nu+\frac52)a^{\frac\nu2+\frac54}}\Big(\Gamma\Big(\frac\nu2+\frac54,az_{\max}^2\Big)-\Gamma\Big(\frac\nu2+\frac54\Big)\Big)\Bigg]\Bigg\},
\label{eq:bdmpszgeneral}
\end{align}
Finally for small $\nu$ behavior of the distribution we expand above equation with $\nu\to0$ and find
\begin{align}
\frac{d\Sigma^{[\nu]}_{q,\med}}{d\chi}&\overset{\nu\to0}{\equiv}-\frac{2}{5}\frac{\alpha_{s,\med}}{\hat{q}L}\sqrt{2\omega_c\omega^5}\Big[z_{\max}^{5/2}\,\Gamma(0,az_{\max}^2)-\frac{1}{a^{5/4}}\nonumber\\
&\times\Big(\Gamma\Big(\frac54,az_{\max}^2\Big)-\Gamma\Big(\frac54\Big)\Big)\Big]+\mathcal{O}(\nu).
\label{eq:bdmpszsmallnu2}
\end{align}
Similar to the previous case, the upper limit of the $z$-integration, i.e., $z_{\max}=\min\{\omega_c/\omega,\,Q_s/(\chi\omega)\}$ divides the medium-induced distribution into two kinematic regimes, separated by an angle $\chi^{\dagger}$. Equating the two regimes, $\omega_c/\omega=Q_s/(\chi^{\dagger}\omega)$, and using $Q_s=\sqrt{\hat{q}L}$ and $\omega_c=\hat{q}L^2/2$, we find
\begin{equation}
\chi^{\dagger}=\frac{Q_s}{\omega_c}=\frac{2}{\sqrt{\hat{q}}\,L^{3/2}},
\end{equation}
which is twice of $\chi^*$ obtained in previous case.  For $\chi<\chi^{\dagger}$, the constraint $\Theta(\omega_c-z\omega)$ dominates and $z_{\max}=\omega_c/\omega$ is independent of $\chi$, whereas for $\chi>\chi^{\dagger}$, the constraint $\Theta(Q_s-|\bfq|)$ takes over leading to $z_{\max}=Q_s/(\chi\omega)$, which goes as $1/\chi$. For the choice of parameters $\hat{q}=1.5$~GeV$^2$fm$^{-1}$ and $L=5$~fm, this gives $\chi^{\dagger}\approx 2.9\times10^{-2}$. Therefore, most of the angular range $\chi\in[10^{-2},1]$ shown in Fig.~\ref{fig:ratiobdmpsz} lies in the $Q_s$ dominated regime, with only the very small values of $\chi$ probing the $\omega_c$ dominated region.

The leading angular scaling of Eq.~\ref{eq:bdmpszsmallnu2} again depends on which of the two regions of $z_{\max}$ dominates. For $\chi<\chi^{\dagger}$, where $z_{\max}=\omega_c/\omega$ is independent of $\chi$, the argument $az_{\max}^2=\chi^2\omega_c^2/(\hat{q}L)\ll1$ over the relevant ranges of $\chi$. Therefore, expanding the incomplete gamma functions in the small $\chi$ limit and combining all the terms give a logarithmic behavior
\begin{equation}
d\Sigma^{[0]}_{q,\med}/d\chi\sim\log\chi+\text{const}.    
\end{equation}
For $\chi>\chi^{\dagger}$, $z_{\max}=Q_s/(\chi\omega)$ and the argument $az_{\max}^2=Q_s^2/(\hat{q}L)\equiv b$ which is $\chi$ independent. Therefore, the $\chi$-dependence in this regime comes from the explicit prefactor $z_{\max}^{5/2}\propto\chi^{-5/2}$ which gives
\begin{align}
&\frac{d\Sigma^{[0]}_{q,\med}}{d\chi}\ =\ C(0)\,\chi^{-5/2},
\end{align}
where the coefficient $C(0)$ is given as
\begin{align}
C(0)&\equiv-\frac{2}{5}\frac{\alpha_{s,\med}}{\hat{q}L}\sqrt{2\omega_c\omega^5}\Big[\Big(\frac{Q_s}{\omega}\Big)^{5/2}\Gamma(0,b)-\Big(\frac{\hat{q}L}{\omega^2}\Big)^{5/4}\nonumber\\
&\qquad\quad\times \Big(\Gamma\Big(\frac54,b\Big)-\Gamma\Big(\frac54\Big)\Big)\Big].
\label{eq:C0}
\end{align}
Since $\hat{q}L/\omega^2=(Q_s/\omega)^2$, both terms in Eq.~\ref{eq:C0} share the common factor $(Q_s/\omega)^{5/2}$. Moreover, this expression gives a power law behavior for all ranges of $\chi$.

Repeating this for $\nu=2$, we find that in the $\omega_c$ dominated regime ($\chi<\chi^{\dagger}$) the leading behavior again remains logarithmic, $d\Sigma^{[2]}_{q,\med}/d\chi\sim\log\chi$. This is because in this regime the leading behavior is dominated by $\Gamma(0,az_{\max}^2)$ which is same for all  $\nu$ values. Moreover, in the $Q_s$ dominated regime ($\chi>\chi^{\dagger}$), the general expression contains two terms that scales as as $\chi^{-7/2}$ and $\chi^{-9/2}$ respectively. We find (numerically) that $\chi^{-7/2}$ term dominates throughout the entire physical range $\chi^{\dagger}\leq\chi\leq1$ due to its large prefactor which increases monotonically from a factor of a few near $\chi^{\dagger}$ to over two orders of magnitude by $\chi\sim1$. Therefore for $\nu=2$, we get
\begin{equation}
\frac{d\Sigma^{[2]}_{q,\med}}{d\chi}\ \simeq\ C(2)\,\chi^{-7/2},
\end{equation}
where the coefficient $C(2)$ reads as
\begin{align}
C(2)&\equiv\frac{4}{7}\frac{\alpha_{s,\med}}{\hat{q}L}\sqrt{2\omega_c\omega^5}\Big(\frac{Q_s}{\omega}\Big)^{7/2}\bigg[\Gamma(0,b)\nonumber\\
&\qquad\quad-\Gamma\Big(\frac74,b\Big)+\Gamma\Big(\frac74\Big)\bigg].
\label{eq:C2}
\end{align}
In the ratio of the two $\nu$ values, i.e., $\nu\to 0$ and $\nu=2$, the $\ln\chi$ dependence present in the $\omega_c$ dominated regime cancels identically. On the other hand, in the $Q_s$ dominated regime the common factor $\alpha_{s,\med}\sqrt{2\omega_c\omega^5}/(\hat{q}L)$ cancels between $C(0)$ and $C(2)$, and the ratio scales as
\begin{equation}
\frac{d\Sigma^{[0]}_{q,\med}/d\chi}{d\Sigma^{[2]}_{q,\med}/d\chi}\ \simeq\ \frac{C(0)}{C(2)}\,\chi\ \equiv\ D\,\chi,
\label{eq:ratioD}
\end{equation}
where $D$ is a constant which depends on medium parameters and energy ($\omega$) of initial quark.  For our default parameters ($\hat{q}=1.5$~GeV$^2$fm$^{-1}$, $L=5$~fm, $\omega=100$~GeV), we find $D\approx -34$. Consequently,  small $\nu$ valued correlators are parametrically enhanced in compared to $\nu=2$ as $\chi$ grows as shown in Fig.~\ref{fig:ratiobdmpsz}.  This tells that small $\nu$ correlators encode richer angular structure than their larger-$\nu$ counterparts. Therefore, varying $\nu$ could provides a tunable angle dependent handle of medium-induced contributions.  Note that while the coefficient $D$ derived above describes the formal $\nu\to0$ limit, evaluating the full expression directly at finite value such as $\nu=0.5$ (used in numerical comparisons) gives a substantially smaller ratio, growing from $|d\Sigma^{[0.5]}_{q,\med}/d\chi|/|d\Sigma^{[2]}_{q,\med}/d\chi|\approx -2.5$ at $\chi=0.3$ to $\approx -4.5$ at $\chi=1$ indicating that the large enhancement captured by $D$ is only realized deep in the $\nu\to0$ limit.

In Fig.~\ref{fig:ratiobdmpsz}, we plot out the ratios (medium only) of $\nu$ correlators with  the standard two point energy correlator as a function of $\chi$, for various values of $\nu\in[0.5,5]$. The band correspond to $\hat{q}=1-2$ GeV$^2$fm$^{-1}$ and $L=5$ fm, with an initial quark energy of $\omega=100$ GeV. For $\nu<1$, we note that the ratios are negative because we have neglected the large contact term which makes the overall distribution positive. Nevertheless, the qualitative trend and shape of the differential distribution ratios are unaffected by this omission as explained below Eq.~\ref{eq:bdmpszdist}. We note that while the larger $\nu$ valued  projected energy correlators ($\nu=3,5$) do not provide much additional information compared to the standard EEC, the smaller $\nu$ valued ones encode richer angular information about medium-induced emissions. 

\subsection{Semi-classical approximation}
\begin{figure}[h]
\centering 
\includegraphics[width=1\linewidth]{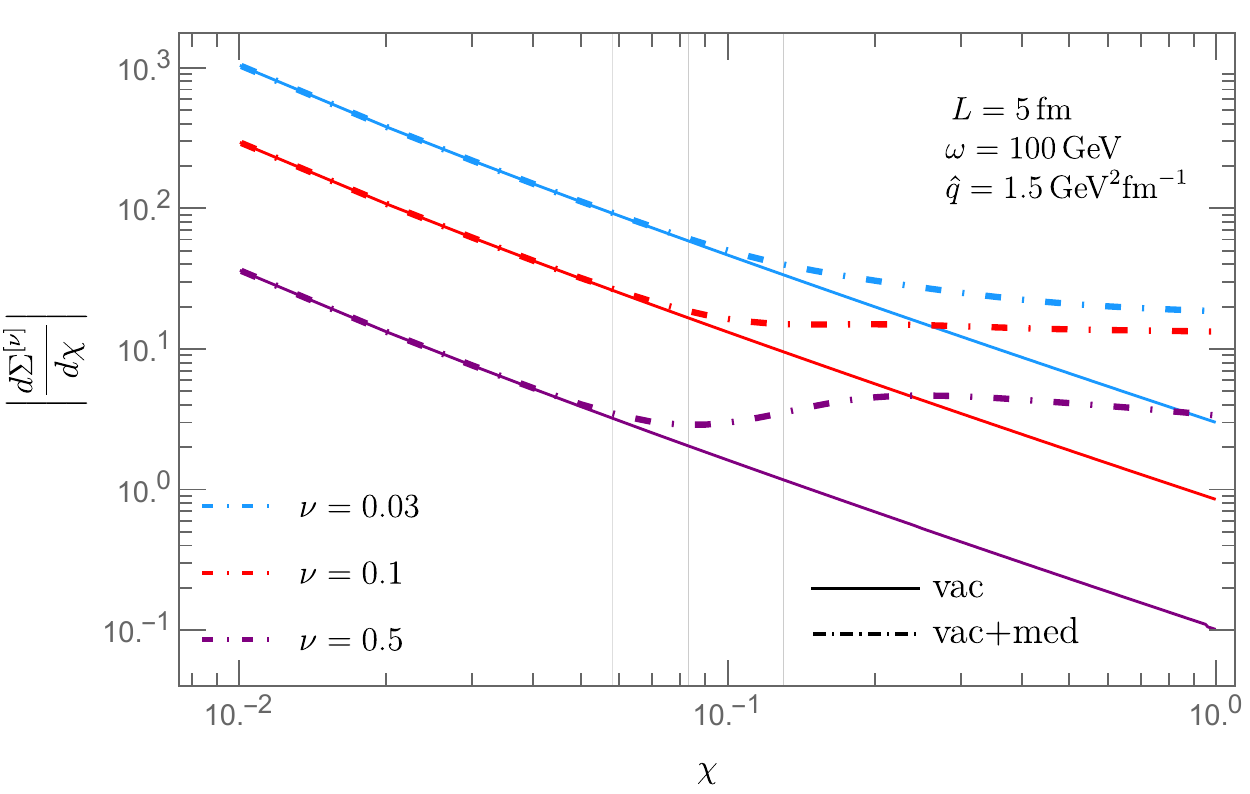}
\caption{ Differential distributions of $\nu$-correlators for $\nu<1$. Solid lines correspond to the vacuum distributions, while dot-dashed lines correspond to the vacuum plus medium-induced distributions given in Eq.~\ref{eq:full}.   } 
\label{fig:miesemi}
\end{figure}

Next, we consider the case of hard medium-induced gluon emissions within the semi-classical approximation. Including the vacuum emissions, the splitting probability is given by~\cite{Dominguez:2019ges,Isaksen:2020npj}
\begin{align}
\frac{d\mathcal{P}_{\rm full}}{dz\,d\theta}&=\frac{d\mathcal{P}_{\rm vac}}{dz\,d\theta}+\frac{d\mathcal{P}_{\rm med}}{dz\,d\theta}\nonumber\\
&=\Big(1+\frac{\alpha_{s,{\rm med}}}{\alpha_s(\mu)}F_{\rm med}(z,\theta)\Big)\frac{d\mathcal{P}_{\rm vac}}{dz\,d\theta}  
\end{align}
where, as before, $\frac{d\mathcal{P}_{\rm vac}}{dz\,d\theta}$ is the vacuum splitting function given in Eq.~\ref{eq:vac}. Here, $\alpha_s(\mu)$ is the running coupling constant appearing in the vacuum cross section and similar to the previous case we take $\alpha_{s,{\rm med}}=0.3$. Moreover, all the medium modifications are described by the function $F_{\rm med}(z,\theta)$ which reads as
\begin{align}
F_{\rm med}(z,\theta)&=\frac{2}{\tau_f}\int_0^L dt\Big[\frac{4}{f_1\hat{q}\tau_f\theta^2 t^2}\Big\{1-e^{-\frac{f_1\hat{q}(L-t)\theta^2 t^2}{4}} \Big\}\nonumber\\
&\cos\Big(\frac{t}{\tau_f}\Big)-\sin\Big(\frac{t}{\tau_f}\Big)\Big]e^{-\frac{1}{12}\hat{q}f_2\theta^2 t^3}   
\label{eq:fmed}
\end{align}
where $\hat{q}$ is jet quenching parameter and $z$ is the energy fraction of the emitted gluon. In Eq.~\ref{eq:fmed} we have omitted non-factorizable terms whose contributions to the energy correlators are expected to be small~\cite{Barata:2023bhh}. The functions $f_1$ and $f_2$ in Eq.~\ref{eq:fmed} are given as
\begin{align}
&f_1=1-2(1-z)+3(1-z)^2\nonumber\\
&f_2=1+(1-z)^2+\frac{2(1-z)}{N_c^2-1},
\end{align}
where $N_c=3$. Moreover, $\tau_f$ is the formation time of the emitted gluon, which is defined as
\begin{equation}
\tau_f=\frac{2}{z(1-z)\omega \theta^2}    
\end{equation}
where $\theta$ is the angle of the emitted gluon from the quark. The full differential distribution is therefore given by
\begin{align}
\frac{d\Sigma^{[\nu]}_{q,{\rm full}}}{d\chi}&= \sum_{i,j\in J}\int dz\, d\theta_{ij}\frac{d\mathcal{P}_{\rm full}}{dz\,d\theta_{ij}}\mathcal{M}^{[\nu]}(z,\theta_{ij},\chi) \nonumber\\
&=\frac{d\Sigma^{[\nu]}_{\rm vac}}{d\chi}+\frac{\alpha_{s,{\rm med}}}{\alpha_s(\mu)}\frac{d\Sigma^{[\nu]}_{\rm med}}{d\chi},
\label{eq:full}
\end{align}
where the second term is obtained using Eq.~\ref{eq:fmed}. The medium induced distribution and their comparison with vacuum ones for various values of $\nu$ is shown in Fig.~\ref{fig:miesemi}. To check the small $\nu$ behavior of the medium-induced distribution in the semi-classical approximation, we first expand the function $F_{\rm med}$ in the $z\to0$ limit and obtain
\begin{align}
F_{\rm med}^{\rm soft}(\theta)=\theta^2\int_0^L dt\, e^{-\frac{3}{16}\hat{q}t^3\theta^2}\Big[\frac{1-e^{-\frac{1}{2}\hat{q}(L-t)t^2\theta^2}}{\hat{q}t^2}+\frac{ t\theta^2}{2} \Big].  
\label{eq:fmedsoft}
\end{align}
Note that $F_{\rm med}^{\rm soft}$ is independent of $z$ in this limit, which allows us to pull it outside the $z$-integration. Using the soft-limit measurement weights (same as in  Eq.~\ref{eq:bdmpszdist}), the medium-induced differential cross section becomes
\begin{align}
\frac{d\Sigma^{[\nu]}_{q,{\rm med}}}{d\chi}\approx\int_0^1 dz\,(\nu z^2-z^{\nu+1})\int\frac{d\theta}{\theta}\,F_{\rm med}^{\rm soft}(\theta)\,\delta(\chi-\theta).
\label{eq:medzint}
\end{align}
Performing the $z$-integration first we get
\begin{align}
\int_0^1 dz\,(\nu z^2-z^{\nu+1})=\frac{\nu}{3}-\frac{1}{\nu+2},
\label{eq:zint}
\end{align}
and then carrying out the angular integration using the delta function we obtain
\begin{align}
\frac{d\Sigma^{[\nu]}_{q,{\rm med}}}{d\chi}\approx\frac{1}{\chi}\Big(\frac{\nu}{3}-\frac{1}{2+\nu}\Big)F_{\rm med}^{\rm soft}(\chi).
\label{eq:medsoft}
\end{align}
Combining Eq.~\ref{eq:medsoft} with the analogous soft-limit vacuum contribution, $d\Sigma^{[\nu]}_{\rm vac}/d\chi\approx(1/\chi)(\nu-1/\nu)$, the resulting medium-induced distribution reads as
\begin{align}
\frac{d\Sigma^{[\nu]}_{q,{\rm full}}}{d\chi}&\overset{\rm soft}{\approx} \frac{1}{\chi} \Bigg[\Big(\nu-\frac{1}{\nu} \Big)+\frac{\alpha_{s,{\rm med}}}{\alpha_s(\mu)}\Big(\frac{\nu}{3}-\frac{1}{2+\nu} \Big) F_{\rm med}^{\rm soft}(\chi) \Bigg]
\label{eq:fmeds}
\end{align}
where, again, the symbol $\approx$ indicates that constant terms are not displayed. Note that similar to the previous two cases of vacuum-like emissions and the BDMPS-Z limit, the medium-induced distributions in the semi-classical approximation also regulate $1/\nu$ behavior, with the corresponding distribution approaching a constant value in the $\nu\to 0$ limit.  Moreover, expanding the bracket in Eq.~\ref{eq:fmedsoft} for $\theta\to0$, we have $\frac{1-e^{-\frac12\hat{q}(L-t)t^2\theta^2}}{\hat{q}t^2}\approx\frac12(L-t)\theta^2$, so that the two terms in the bracket combine as
\begin{equation}
\frac12(L-t)\theta^2+\frac{t\theta^2}{2}=\frac{L\theta^2}{2},
\end{equation}
which is independent of $t$. The exponential outside the bracket approaches unity as $\theta\to0$, therefore the remaining $t$-integral in small angle limit leads to
\begin{equation}
F_{\rm med}^{\rm soft}(\chi)\ \overset{\chi\to0}{\longrightarrow}\ \frac{L^2}{2}\,\chi^4+\mathcal{O}(\chi^6),
\label{eq:fmedsmallchi}
\end{equation}
which is same for all $\nu$ values. Substituting Eq.~\ref{eq:fmedsmallchi} into the medium-induced term of Eq.~\ref{eq:fmeds}, we find that this contribution scales as $\chi^{-1}F_{\rm med}^{\rm soft}(\chi)\sim\chi^3$. The semi-classical medium correction is thus parametrically suppressed relative to the vacuum contribution ($\sim 1/\chi$) at small angles. 
\begin{figure}[t]
\centering 
\includegraphics[width=1\linewidth]{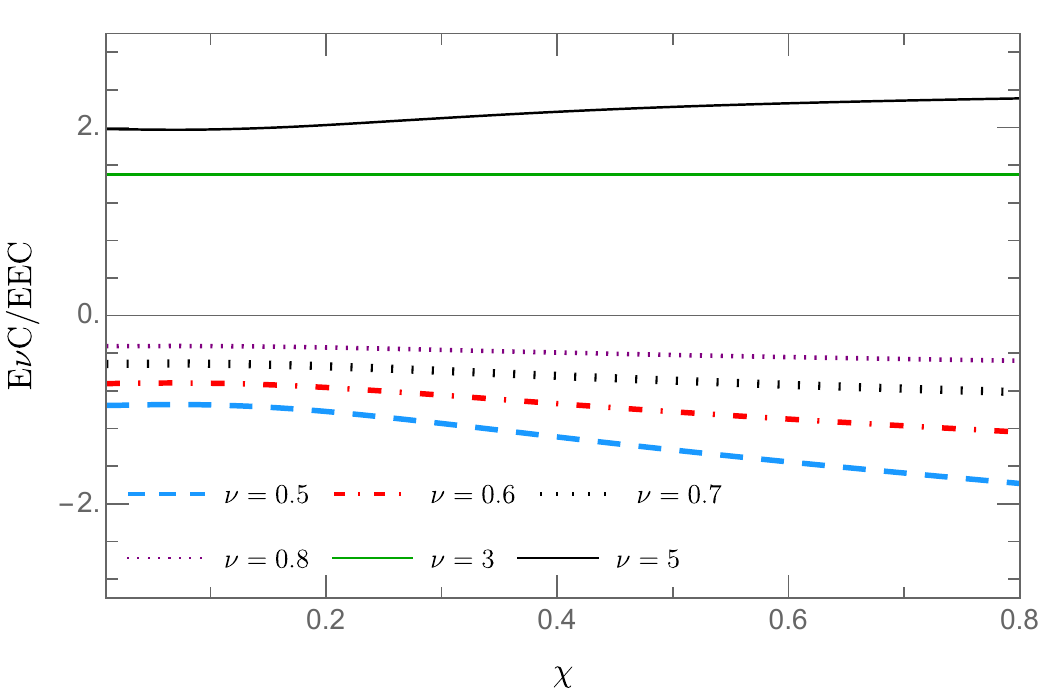}
\caption{Ratios of the medium-only $\nu$-correlators to the standard two point energy correlator, for $\nu$ ranging from $0.5$ to $0.8$, as a function of the angular separation $\chi$. The distributions are obtained using the BDMPS-Z spectrum given in Eq.~\ref{eq:bdmpszdist}, for an initial quark energy $\omega=100$ GeV and jet quenching parameter $\hat{q}=1.5$ GeV$^2$fm$^{-1}$. The medium parameters are set to $L=5$ fm and $\alpha_{s,\med}=0.3$.} 
\label{fig:ratiosemi}
\end{figure}

In Fig.~\ref{fig:ratiosemi}, using the full structure of $F_{\rm med}$, we plot the ratios of the medium-induced projected energy correlator distributions to the standard two-point energy correlator for various values of $\nu$. The medium parameters are $\hat{q}=1.5$~GeV$^2$fm$^{-1}$ and $L=5$~fm, the same as those used in Fig.~\ref{fig:ratiobdmpsz}. We note that the ratios are approximately flat in the small angular region ($\chi<0.2)$ which is different from the one observed in BDMPS-Z case (linear, see Eq.~\ref{eq:ratioD}). This can be understood from the soft-limit result of Eq.~\ref{eq:medsoft}: since $F_{\rm med}^{\rm soft}(\chi)$ is same for all $\nu$ values, it cancels in the ratio to the standard EEC, leaving the magnitude of the ratio fixed by the $\chi$-independent factor $\big(\tfrac{\nu}{3}-\tfrac{1}{2+\nu}\big)$. The mild $\chi$-dependence visible at larger $\chi$ originates from subleading, finite-$z$ corrections to $F_{\rm med}(z,\theta)$ beyond the strict soft limit of Eq.~\ref{eq:fmedsmallchi}, so that the overall qualitative behavior of the ratios in Figs.~\ref{fig:ratiobdmpsz} and~\ref{fig:ratiosemi} is captured by small $\nu$ and small $\chi$ expansions of the medium-induced distributions derived above. The comparison between the medium-induced and vacuum distributions, on the other hand, is governed by the $1/\nu$ enhancement present in the vacuum distribution as $\nu\to0$, which as shown above is absent in both medium-induced distributions.

In Fig.~\ref{fig:miesemi}, we plot the vacuum and medium-induced distributions given in Eq.~\ref{eq:full} for $\nu<1$. The solid line represents the vacuum distribution and the dot-dashed line represents the full distribution from Eq.~\ref{eq:full}. To improve the visibility of the plot, we have multiplied the differential distribution by $-1$ and displayed it on logarithmic scale. For vacuum distribution, we use the one-loop running coupling constant and set the scale to $\mu=\omega\chi$, while for the medium distribution we use a fixed value, $\alpha_{s,\mathrm{med}}=0.3$. The remaining medium parameters are $L=5$~fm, $\hat{q}=1.5$~GeV$^2$fm$^{-1}$, and $\omega=100$~GeV. At small $\chi$, the vacuum and full distributions coincide closely for all three values of $\nu$ shown, consistent with our earlier finding (see Eq.~\ref{eq:fmedsmallchi})  that the medium-induced contribution is parametrically suppressed relative to vacuum in this region. As $\chi$ increases past the vertical grey lines -- which mark the onset of the medium-induced enhancement for each $\nu$ -- the two curves visibly separate. We note that as $\nu$ approaches smaller values, this large-angle enhancement due to medium-induced emissions is suppressed. This is particularly due to the $1/\nu$ behavior of vacuum distributions which grows as $\nu\to 0$ while the medium induced distribution approaches a constant as shown in both BDMPS-Z and semi-classical approximations.~\footnote{It is worth clarifying an apparent behavior in the small $\nu$ distributions described above. When comparing medium-induced distributions to one another across different values of $\nu$, we find that the small $\nu$ valued correlators are enhanced compared to $\nu=2$, as encoded by the factor $D\chi$ in Eq.~\ref{eq:ratioD}. However, when comparing the same medium-induced contribution to the vacuum distribution at fixed $\nu$, the small $\nu$ region is suppressed because vacuum distribution diverges as $1/\nu$ while the medium-induced terms approaches a finite constant as $\nu\to0$.} We have also verified this observation using the harmonic oscillator approximation~\cite{Isaksen:2020npj}.  It is worth mentioning that the semi-classical approximation yields a somewhat large enhancement in the large-angle region~\cite{Barata:2023bhh}, and this enhancement is known to be sensitive to the treatment of the underlying splitting kernel employed here. In the future, it would be useful to incorporate the more accurate splitting functions derived in Refs.~\cite{Leitao:2026fgh,Andres:2026qrt}. Furthermore, another effect not accounted for here is energy loss, which would further suppress the correlator distribution in the large-angle region for smaller $\nu$-values. In this case, we expect the small $\nu$ values of the generalized energy correlators to be more sensitive to medium response. We explicitly check this simulated events with JEWEL.

\section{JEWEL Simulations~\label{sec:jewelresults}}
In this section, we discuss the phenomenological relevance of small $\nu$-correlators using inclusive jet samples generated with the Monte Carlo event generator JEWEL. For  samples, we select jets with $p^J_T=100$--$120$ GeV, radius $R=0.4$, and rapidity $|\eta^{J}|<2$. To analyze $\nu$-correlators for various values of $\nu$, we use the method developed in Ref.~\cite{Alipour-fard:2024szj}. For the medium-modified $\nu$-correlator distributions with recoil switched on in JEWEL, we additionally apply the constituent subtraction method for background subtraction, as discussed in Ref.~\cite{KunnawalkamElayavalli:2017hxo}. Unless explicitly stated for all plots use the default JEWEL settings and a centrality class of $0$--$10\%$. Finally, we use the {\textsc{RIVET v3.1.7}} analysis framework~\cite{Buckley:2010ar} throughout in this section.    
  
\begin{figure}[h]
\centering 
\includegraphics[width=1\linewidth]{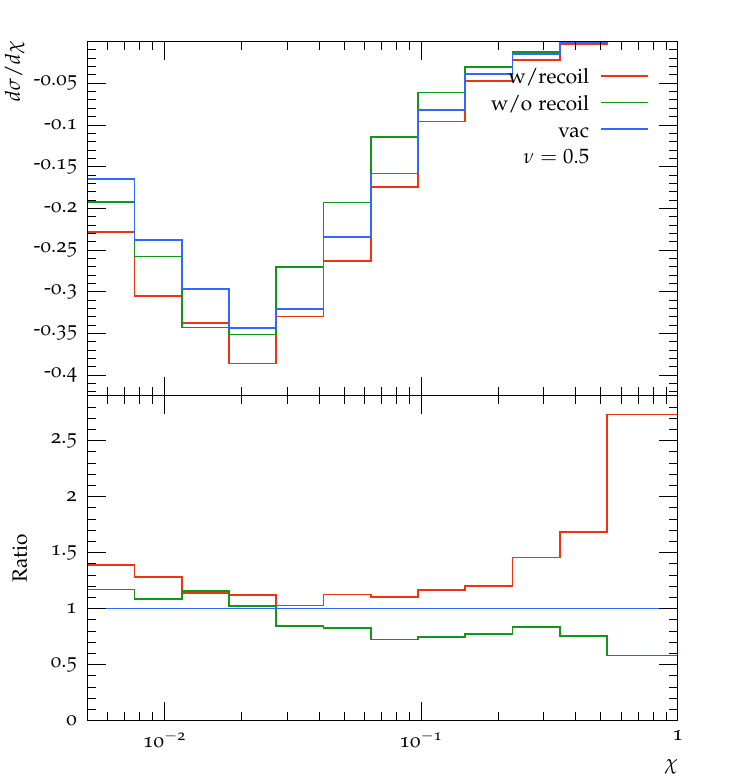}
\caption{ Differential distribution as a function of the angle $\chi$ for $\nu=0.5$, with jet parameters $R=0.4$, $p^{J}_T\in [100,120]$ GeV, and $|\eta^J|<2$. The distributions are scaled by a constant factor of $100$. Ratios are taken relative to the vacuum distribution, represented by the blue line.} 
\label{fig:incnu0p5}
\end{figure}

In Fig.~\ref{fig:incnu0p5}, we show the distributions for $\nu=0.5$ for both the vacuum and medium cases, generated using the process $PP\to JJ$. The jet parameters are the same as those mentioned earlier, and for the medium we set the initial temperature $T=0.45$ GeV, centrality $0$--$10\%$, and use the default setting for the Debye screening mass\footnote{We have checked that increasing the Debye mass in JEWEL enhances the distribution in all angular regions by a constant factor for $\nu<1$.}. The blue curve is vacuum distribution, while the green and red curves represent the medium-induced distributions with recoil switched off and switched on in JEWEL, respectively. 

We first note that the differential distribution for $\nu<1$ is negative across the full range of $\chi$, for the same reason as explained below Eq.~\ref{eq:bdmpszdist}. To improve visibility, we scale all vacuum and medium distributions in Fig.~\ref{fig:incnu0p5} by a constant factor of $100$; this scaling does not affect the shape of the ratios of medium to vacuum distributions. The medium-induced distribution with recoil switched off (green) is a little bit enhanced  in the small angle region and somewhat suppressed in the large angle region. In contrast, with recoil switched on (red), the distribution is enhanced across the entire angular range considered in Fig.~\ref{fig:incnu0p5}. This is seen explicitly in the ratios of the medium to vacuum distributions shown in the lower panel of the same figure. The horizontal blue line is vacuum reference line at unity. The red line is the ratio of the medium distribution with recoil on to the vacuum distribution. Here, while the small rise at smaller angles is due to energy loss, the large angle enhancement arises from recoil, since the medium-induced emissions alone (green line) are suppressed there, as also observed in Fig.~\ref{fig:miesemi}. Since medium-induced contributions approach a constant value as $\nu\to0$ while the vacuum distribution diverges as $1/\nu$, the analytic calculation predicts that medium-induced gluon emissions should yield small contribution in the large angle region for small $\nu$. 
\begin{figure}[h]
\centering 
\includegraphics[width=1\linewidth]{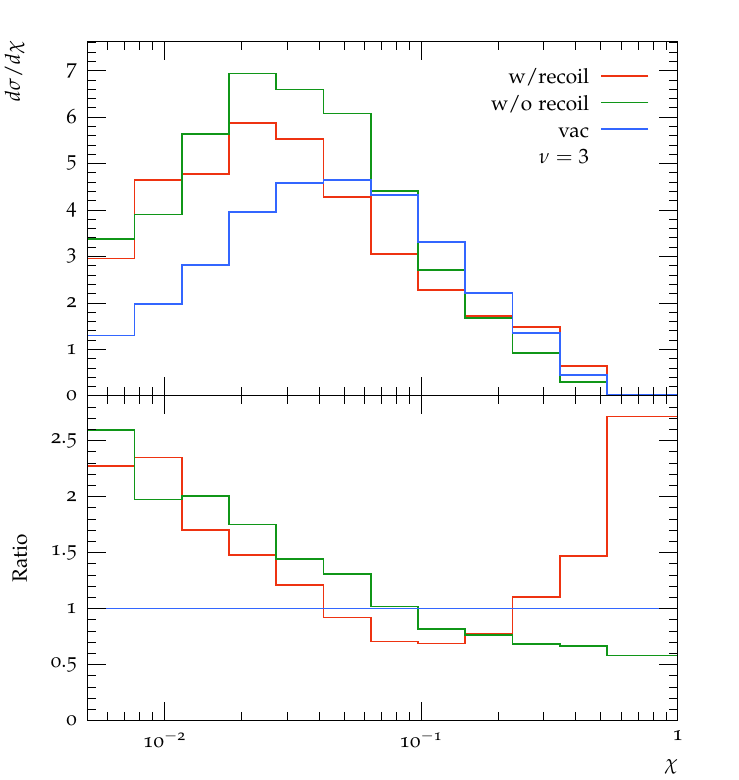}
\caption{ Differential distribution as a function of the angle $\chi$ for $\nu=3$, with jet parameters $R=0.4$, $p^{J}_T\in [100,120]$ GeV, and $|\eta^J|<2$. The distributions are scaled by a constant factor of $100$. Ratios are taken relative to the vacuum distribution, represented by the green line.} 
\label{fig:incnu3}
\end{figure}
It is worth mentioning that the constant shift from the vacuum baseline in the intermediate region would come closer to one after proper normalization of the distributions. However, this normalization is not expected to change the shape of the ratios between the distributions. 

To compare the small $\nu$ projected energy correlators with higher point projected energy correlator, in Fig.~\ref{fig:incnu3}, we show vacuum and medium induced distributions for $\nu=3$. Here the distributions are neither normalized nor scaled by any constant factor. Comparing the ratios for small and large $\nu$ values, we see that while the large angle enhancement appears to start around the similar angular regions as in Fig.~\ref{fig:incnu0p5}, the small angle region is affected differently compared to $\nu=0.5$. Further these two figures with the one in Fig.~\ref{fig:incnu2} (for $\nu=2$) we note that ratio of medium and vacuum distributions $\nu=2$ lies somewhere between $\nu=0.5$ and $\nu=3$.  This seems to suggest that a scan of projected energy correlators from small to larger $\nu$ values may help to understand medium response, energy loss, and medium-induced radiation contributions to jet modifications in HICs.

\section{Summary}
\label{sec:summary}
In this work, we study the behavior of projected energy correlators, with particular focus on $\nu<1$, and show that a scan over projected energy correlators can be used to quantify various effects in jet-medium interaction dynamics. Starting with vacuum-like emissions, which have shorter formation times than medium-induced emissions, i.e., $\tau_{f,\vac}\ll \tau_{f,\med}$, we showed that the small $\nu$ enhancement present in the vacuum distribution is regulated by formation-time constraints. As a result, the small $\nu$ region for medium-induced emissions saturates at a constant value as $\nu\to 0$, indicating that in-medium emissions are suppressed relative to vacuum at smaller values of $\nu$ for projected energy correlators. 

Next, we examined the projected energy correlator distributions for medium-induced emissions, accounting for multiple scatterings within both the BDMPS-Z and semi-classical approximations. We showed that compared to EEC, correlators with smaller $\nu$ value encode richer angular information about medium-induced emissions than their higher-$\nu$ counterparts and this behavior is qualitatively similar in both approximations. Furthermore, we found that multiple interactions between the jet and the medium regulate the $1/\nu$ enhancement seen in the vacuum distribution. We confirmed this behavior for both soft and relatively harder medium-induced emissions. Finally, compared to vacuum baseline the medium-induced contributions are suppressed at smaller $\nu$-values, as shown in Fig.~\ref{fig:miesemi}.

We observed similar behavior for $\nu$-correlators in Fig.~\ref{fig:incnu0p5} for $\nu=0.5$ using simulated events from JEWEL. While the small enhancement in the small angle region can be attributed to energy loss, the large angle enhancement arises from recoil within the JEWEL framework. Morevoer, since the medium-induced contributions in the intermediate region are shifted by an approximately constant factor, their effect is expected to cancel in the normalized distributions. As a result, for $\nu = 0.5$, the observable appears to receive enhanced contribution from recoil within the JEWEL framework. This is consistent with our analytic expectation from Sec.~\ref{sec:meddist} that medium-induced emissions alone are suppressed at large angle for small $\nu$, so that observed enhancement in this region is expected originate from medium response. A more detailed investigation, including the medium response as incorporated in Ref.~\cite{Barata:2024ukm} will be carried out in future work to identify the range of $\nu$ values that could be used to disentangle various medium-induced effects.  

Comparing Figs.~\ref{fig:incnu0p5}, \ref{fig:incnu3}, and \ref{fig:incnu2}, we further note that while the large angle enhancement appears qualitatively similar across $\nu=0.5$, $2$, and $3$, the small angle enhancement differs drastically between these values. This indicates that a scan over both larger- and smaller-valued $\nu$ projected energy correlators could be pivotal in isolating distinct effects such as medium response, medium-induced emissions, and energy loss. While the computations in this paper are restricted to leading order, it is worth noting that higher-order theoretical calculations will provide deeper insight and better accuracy in understanding jet-medium interaction dynamics in heavy-ion collision environments.

\section{\label{sec:acknowledgments}Acknowledgments}
B.S. would like to thank Raghav Kunnawalkam Elayavalli, Varun Vaidya, João Barata, Andrey V. Sadofyev, Liliana Apolin\'ario and Ankita Budhraja for helpful discussions. This work  is supported by Fundação para a Ciência e a Tecnologia (FCT) through the ERC-PT A-Projects ‘Unveiling’, financed by PRR, NextGenerationEU and Fundação para a Ciência e a Tecnologia (FCT) under con-
tracts2023.15319.PEX(https://doi.org/10.54499/2023. 15319.PEX).

\appendix

\section{Additional $\nu$-correlator distributions}
\label{app:mult}

\begin{figure}[h]
\centering 
\includegraphics[width=1\linewidth]{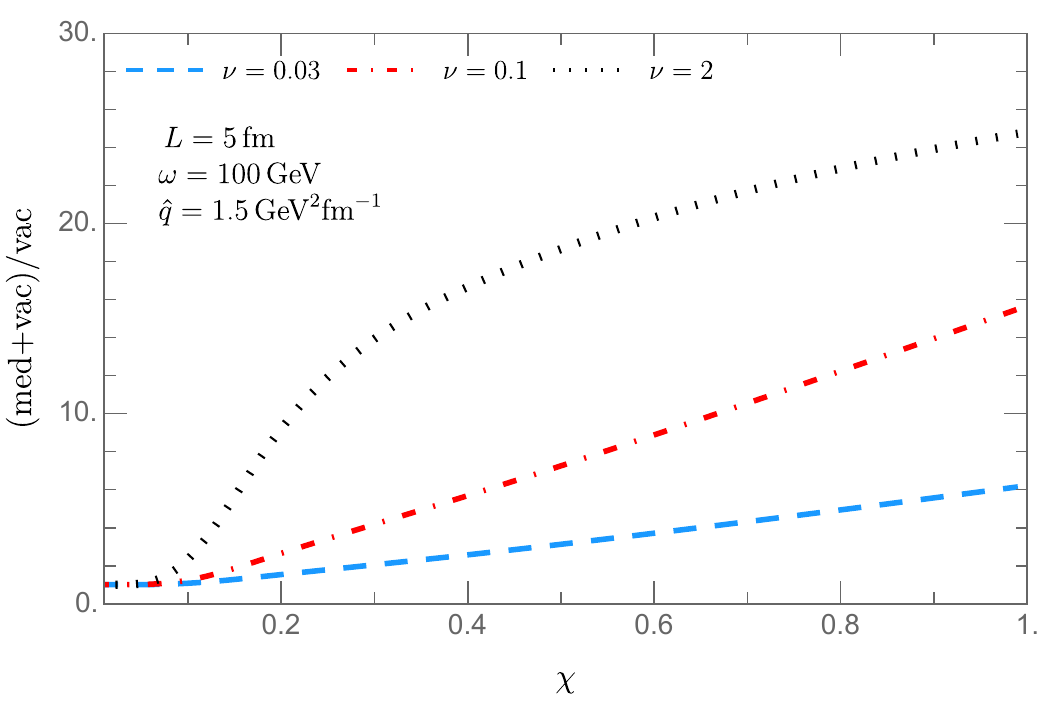}
\caption{Ratios of the full (vacuum + medium-induced) $\nu$-correlator distributions to the corresponding vacuum distributions as a function of $\chi$ for various values of $\nu=0.03,0.1,2$ within semi-classical approximation of multiple scatterings of energetic parton and medium. } 
\label{fig:vacratio}
\end{figure} 

In Fig.~\ref{fig:vacratio}, we show the ratios of the $\nu$-correlator (vacuum $+$ medium) distributions to the vacuum distribution for $\nu$ values considered in Fig.~\ref{fig:ratiosemi}, i.e., $\nu=0.03$, $0.1$, and $2$. 

\begin{figure}[h]
\centering 
\includegraphics[width=0.8\linewidth]{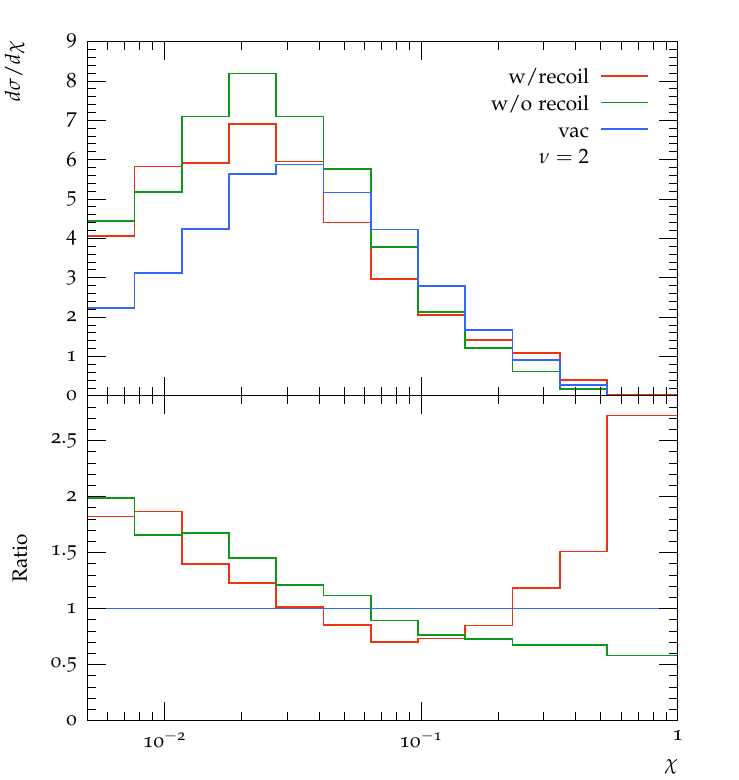}
\caption{ Differential distribution of projected energy correlators as a function of the angle $\chi$ for $\nu=2$, with jet parameters $R=0.4$, $p^{J}_T\in [100,120]$ GeV, and $|\eta^J|<2$. In the lower paned the ratios are taken relative to the vacuum distribution which represented by the blue line. The medium parameters are same as the one considered in Fig.~\ref{fig:incnu3}.} 
\label{fig:incnu2}
\end{figure}

For the energetic parton and its medium multiple scatterings with the medium, we use the full expression for $F_{\rm med}$ given in Eq.~\ref{eq:fmed}. The medium parameters and the initial quark energy are the same as those used in Fig.~\ref{fig:miesemi}. As noted earlier in the discussion of Fig.~\ref{fig:ratiosemi}, the medium-induced contribution is suppressed as we move toward smaller values of $\nu$. This behavior is consistent with our earlier observation that vacuum emissions dominate the full distribution in the small $\nu$ limit, since medium-induced contributions remain finite while the vacuum one grows as $\nu\to 0$. Consequently, the ratios shown in Fig.~\ref{fig:vacratio} decrease systematically as $\nu$ decreases.

In Fig.~\ref{fig:incnu2}, we show the distribution of the standard two point energy correlator, i.e., $\nu=2$, for the vacuum case as well as the recoil-on (red curve) and recoil-off (green curve) scenarios in JEWEL. The lower panel displays the ratios of the medium-induced distributions to the vacuum distribution. The medium and jet parameters are the same as those used in Figs.~\ref{fig:incnu0p5} for $\nu=0.5$ and \ref{fig:incnu3} for $\nu=3$. 

\bibliography{main}
\end{document}